\documentclass[apj]{emulateapj}

\usepackage{graphics}
\usepackage{epsfig}

\shorttitle{Distance to SNR G349.7+0.2}
\shortauthors{Tian \& Leahy}

\begin{document}

\title{SNR G349.7+0.2: A $\gamma$-ray source in the far 3 kpc arm of the Galactic center}

\author{W.W. Tian\altaffilmark{1,2}, D.A. Leahy\altaffilmark{2}}
\altaffiltext{1}{National Astronomical Observatories, CAS, Beijing 100012, China. tww@bao.ac.cn}
\altaffiltext{2}{Department of Physics \& Astronomy, University of Calgary, Calgary, Alberta T2N 1N4, Canada}
 
\begin{abstract}
We analyze the HI absorption profile for TeV Supernova Remnant (SNR) G349.7+0.2 based on updated knowledge of the inner Galaxy's structure. We significantly revise its kinematic distance from the previous
 $\sim$ 22 kpc to $\sim$ 11.5 kpc, indicating it is in the far 3 kpc arm of the Galactic center. We give a revised age of $\sim$ 1800 year for G349.7+0.2 which has a low explosion energy of $\sim$ 2.5 $\times$10$^{50}$ ergs. This removes G349.7+0.2 from the set of brightest SNRs in radio and X-ray to $\gamma$-ray wavebands and helps us to better understand $\gamma$-ray emission originating from this remnant. However, one needs to use caution when discussing old kinematic distances of Galactic objects (e.g., SNRs, pulsars and HII regions) in the range of -12$^{0} \le l \le 12^{0}$ with distance estimates of $\ge$ 5.5 kpc.  
\end{abstract}


\section{Introduction and Data}

It is widely believed that most Galactic cosmic rays probably 
originate from supernova remnants (SNRs). Recently the TeV $\gamma$ -ray emission from SNRs has attracted interest because of new TeV observations that have identified some sources with SNRs (http://tevcat.uchicago.edu/ and http://www.mpi-hd.mpg.de/hfm/HESS/pages/home/sources/). Most of the TeV SNRs detected so far are within short distances (e.g. Cas A: 3.4 kpc (Krause et al. 2008), RC86: 2.5 kpc (Bamba et al. 2000), IC443: 1.5 kpc (Fesen 1984), 
SN1006: 2.2 kpc (Winkler et al. 2003), G353.6-0.7: 3.2 kpc (Tian et al. 2008)).
Tian \& Zhang (2013) listed 61 TeV $\gamma$ -ray sources which were likely identified as SNRs/PWNe, and showed that most of them are less than 6 kpc away. 

Many SNRs' kinematic distances are obtained by observing radial velocities of atomic and molecular lines toward the SNRs and using a circular rotation curve model. 
A circular rotation curve model is generally not suitable for Galactic center (GC) objects because the gas likely follows oval orbits in the inner Galaxy, i.e. inside the 3 kpc ring. Many studies have shown that it is extremely difficult to determine precise distances along the line-of-sight to objects in the 3 kpc ring. In this case, previous kinematic distance estimates to some remnants might contain a large uncertainty or even be incorrect. Recently-finished radio and infrared surveys reveal detailed kinematic images of gas and dust in GC region (Dame \& Thaddeus 2008, McClure-Griffiths et al. 2012). Several good gas flow models in the GC have been presented based on recent observations (e.g. Rodriguez-Fernandez \& Combes 2008). 

In this letter, we study a newly detected TeV source: SNR G349.7+0.2 (Trichard et al. 2013), which has long been believed to be at a large distance of $\sim$ 22 kpc away, meaning that it resides on the other side of the Milky Way, based on observations of HI, 1720 MHz OH masers, CO and H$_{2}$CO etc. (Caswell et al. 1975, Frail et al. 1996, Reynoso \& Mangum 2001). A high resolution radio continuum image of G349.7+0.2 at 18 cm was presented previously by Lazendic et al. (2010).
We construct an HI absorption line profile and utilize new studies of the inner Galactic structure in order to constrain the distance and other key physical parameters of SNR G349.7+0.2. The 1420 MHz radio continuum and 
21 cm HI emission data sets are from the Southern Galactic Plane Survey (SGPS),using the Australia Telescope Compact Array and the Parkes 64m single dish telescope (Haverkorn et al. 2006). The continuum observations have a resolution of 100~arcsec and a sensitivity better than 1~mJy/beam. The HI data have an angular resolution of 2~arcmin, an rms sensitivity of $\sim$1~K and a velocity resolution of 1~km~s$^{-1}$. The CO ($J$=1-0) data for G349.7+0.2 is from survey data from the 12 m NRAO telescope (see Reynoso \& Mangum 2000 for details).

\section{Analysis and Results}
\subsection{Spectra}
Methods for extracting HI absorption spectrum have been introduced in earlier papers (Tian et al. 2007; Leahy \& Tian 2008; Leahy \& Tian 2010). 
The HI emission spectrum of SNR G349.7+0.2 is shown in the upper panel of Fig. 1. IN this spectrum, there are several HI emission peaks from -113 km s$^{-1}$ to 35 km s$^{-1}$. 
The HI absorption and CO emission spectra are shown in the lower panel of Fig. 1. 
SNR G349.7+0.2 is strong radio source, which results in a strong HI absorption signal.
An estimate of the error in exp(-$\tau(v)$)  of Fig. 1 is the r.m.s. for velocity channels with no emission ($>$+50 km/s), which is 0.0063.
Artifacts in the continuum image cause another type of error. We have analyzed absorption spectra both with the same source region and with different backgrounds. We find that the artifacts only affect the normalization and not the shape of the HI absorption spectra. We find that the error in normalization is about 20\% in depth of the HI absorption.  
We find that each CO emission peak has an associated HI absorption feature except for the 16.5 km s$^{-1}$CO peak. The highest positive velocity absorption feature is at 6$\pm$7 km s$^{-1}$. The lowest negative velocity absorption feature is likely at -110$\pm$10 km s$^{-1}$ (a weak absorption feature appears at -189$\pm$2 km s$^{-1}$ which is real, see section 3.1).
We calculate the absorption column density from the HI absorption spectrum, N$_{HI}$ $\sim$ 1.44 $\times$ 10$^{22}$ cm$^{-2}$, taking a value of T$_{s}$= 100 K (using N$_{HI}$= 1.9 $\times$ 10$^{18} \int\tau d{\it{v}} T_{s}$ cm$^{-2}$, Dickey \& Lockman 1990).     

\begin{figure} 
\vspace{60mm} 
\begin{picture}(80,80)
\put(-10,0){\includegraphics{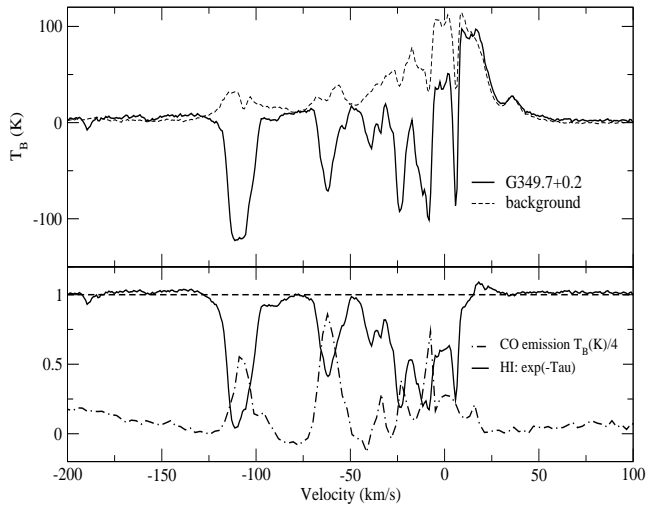}}
\end{picture}
\caption{HI and $^{12}$CO spectra of G349.7+0.2}
\end{figure} 

\subsection{HI Channel maps}

\begin{figure*}
\vspace{50mm}
\begin{picture}(80,80)
\put(-35,235){\includegraphics{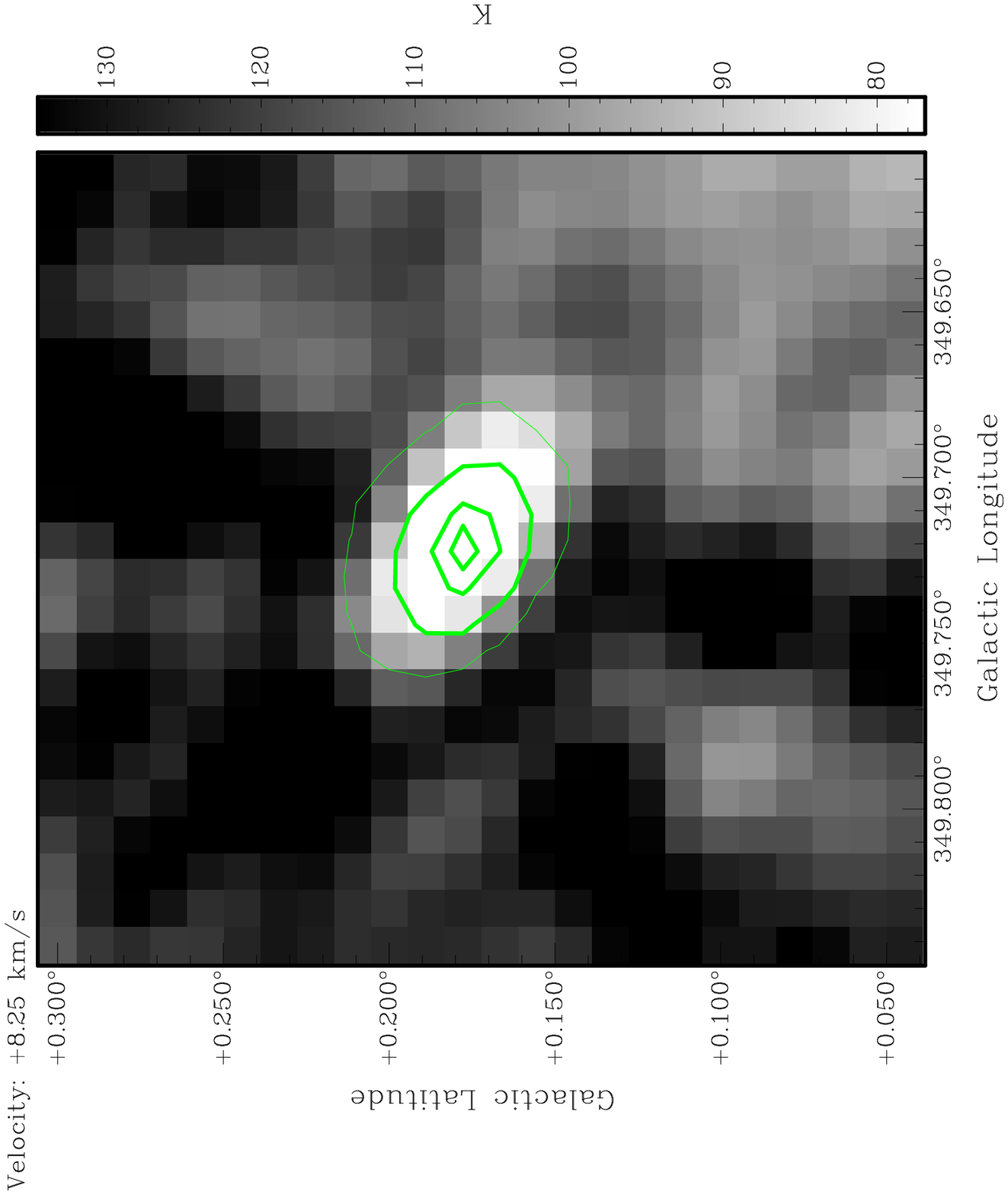}}
\put(220,235){\includegraphics{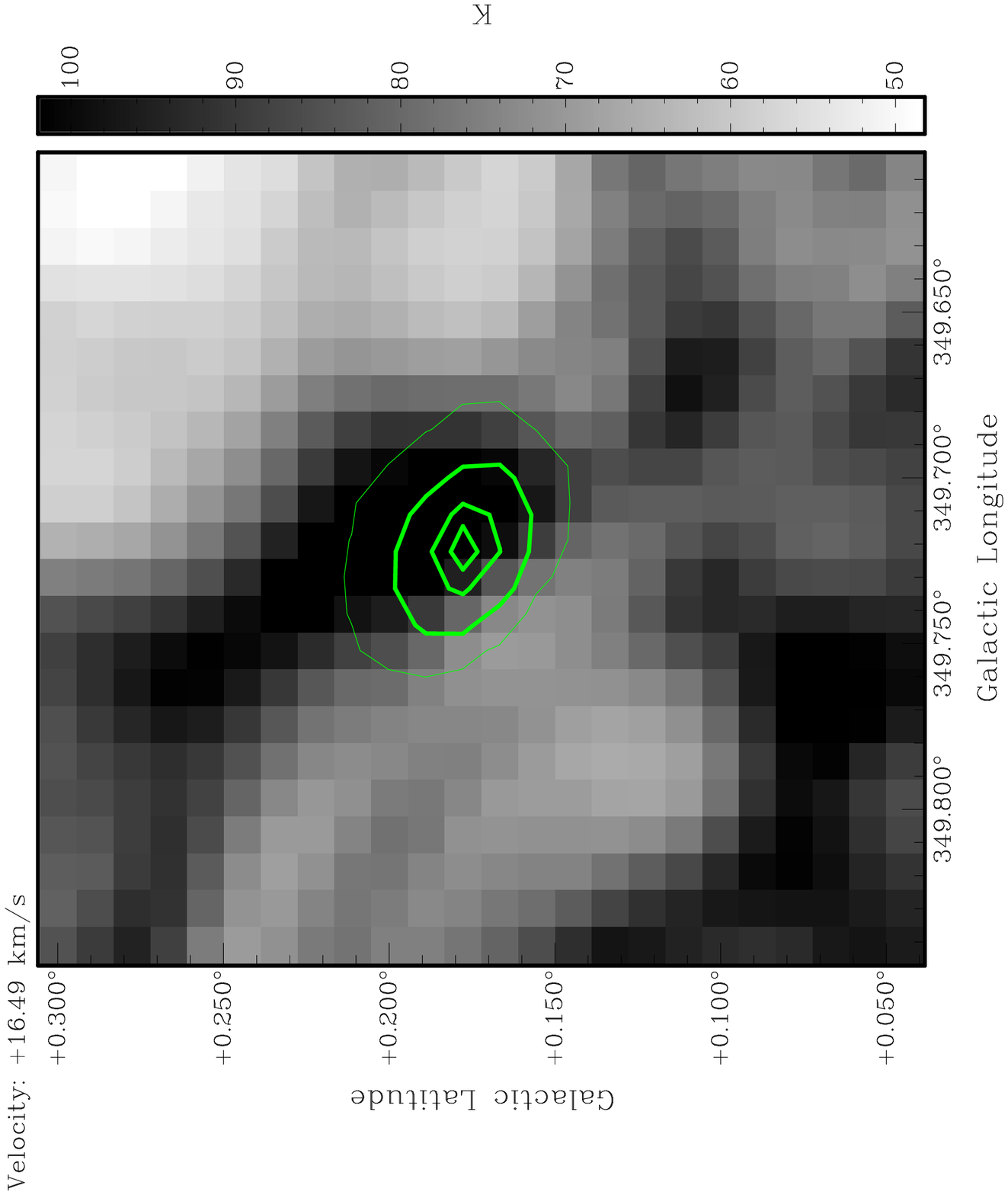}}
\end{picture}
\caption{HI channel maps at velocities of 8.3 and 16.5 km/s. The contours at 0.2, 2, 6, 8 Jy/Beam are from the 1420 MHz continuum image of SNR G349.7+0.2.}
\end{figure*}

\begin{figure}
\vspace{60mm}
\begin{picture}(80,80)
\put(10,-15){\includegraphics{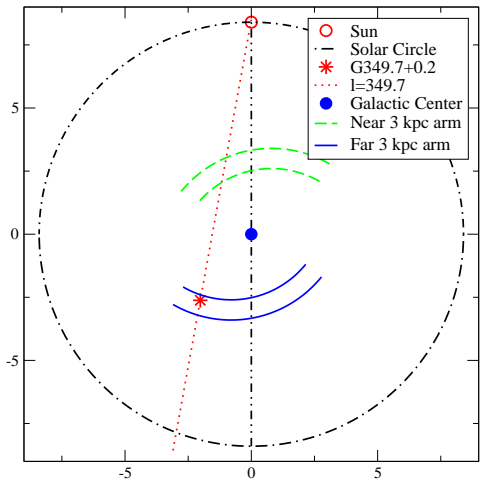}}
\end{picture}
\caption{Diagram indicating the location of SNR G349.7+0.2 in the far 3 kpc arm of the Galactic center. The Galactic center structure is referenced in to Churchwell et al. (2009)}

\end{figure}

We use the HI channel maps to confirm the reality of all above-mentioned absorption features, including the 189 km s$^{-1}$ absorption feature. Absorption continues to appear from the channel map at velocity  $\sim$-123 km s$^{-1}$s through the map at velocity  $\sim$13 km s$^{-1}$;. Fig. 2 shows the HI channel maps at 8.3 km s$^{-1}$ and 16.5 km s$^{-1}$  towards the SNR. The left panel of Fig. 2 shows clear absorption against the SNR. The right one reveals that the remnant sits on a bright HI patch which is associated with the 16.5 km s$^{-1}$ CO cloud. This is consistent with above absorption spectrum analysis, i.e. no HI absorption associated with the HI and CO clouds at 16.5 km s$^{-1}$. This directly confirms that the 16.5 km s$^{-1}$ clouds are behind the remnant. 

\section{Discussion and Conclusion} 

HI absorption spectra towards Galactic radio-bright objects have been used to estimate the line-of-sight distances based on a rotation curve model. 
However, it is a challenging job to constrain the distances to Galactic objects in the inner 3 kpc of the Galaxy because the objects likely follow non-circular orbits in the region (see Weiner \& Sellwood 1999). 

The situation has improved due to recent studies of the Galactic center's structure.
High resolution HI observations (McClure-Griffiths et al. 2012) of the inner Galaxy show greater details of the region's structure. 
CO observations (Dame \& Thaddeus 2008) have detected the symmetric expanding 3 kpc arms, supporting the existence of a bar at the center of our Galaxy. 
Rodriguez-Fernandez \& Combes (2008) iroveprovedd a gas distribution model for the GC region by simulating the gas dynamics in the inner Galaxy. 
This model gives us an opportunity to better understand the HI and CO spectra and refine the distance estimation of G349.7+0.2. 
We adopt the 1985 IAU standard for Galactic parameters in this paper, i.e. R$_{o}=$ 8.5 kpc (the distance between Sun and the GC) and V$_{o}$ = 220 km/s (circular velocity at R$_{o}=$).  

\subsection{Distance of G349.7+0.2}

The HI terminal velocity at $l$=349$^{0}$ to 350$^{0}$ is about -190 km~s$^{-1}$ (see Fig. 2 of Weiner \& Sellwood 1999 and of Burton \& Liszt 1993). 
We detect a weak absorption feature at -189 km~s$^{-1}$ in the spectrum of SNR G349.7+0.2, which is likely caused by clumps of HI possibly accelerated by a SNR or HII region along this line-of-sight. 
The lowest-velocity major absorption feature, except for the -189 km~s$^{-1}$ feature, is at $\sim$ -110 km~ s$^{-1}$ in the SNR spectrum (Fig. 2). 
This velocity is consistent with the expected radial velocity (101.6$\pm$8.7 km~ s$^{-1}$) of the near 3 kpc arm in the line-of-sight to $l$=349$^{0}$ to 350$^{0}$ (equation (3) of Jones et al. 2013, see also Fig. 1 of Dame \& Thaddeus 2008). 
This is convincing evidence that the remnant is beyond the near 3 kpc arm. 

In addition, the highest-velocity absorption in the spectrum of G349.7+0.2 is at 6$\pm$7 km ~s$^{-1}$. 
This is consistent with the expected radial velocity (16.3$\pm$15.6 km~s$^{-1}$) of the far 3 kpc arm toward $l$ $\sim$-10.3$^{0}$ (Fig. 1 and Equation (4) of Jones et al. 2013). 
However, in the SNR spectrum we found no HI absorption associated with the HI and CO emission clouds at 16.5 km ~s$^{-1}$, which exactly lie in the far 3 kpc arm extending at least 20$^{\circ }$ in longitude, starting at l=-12$^{\circ}$ (Dame \& Thaddeus 2008). 
These two pieces of evidence argue that the remnant is in the far 3 kpc arm; the cold HI gas at  6$\pm$7 km ~s$^{-1}$ is in front of the remnant, and the HI and CO clouds at 16.5 km ~s$^{-1}$ are behind the remnant (see Fig.3).

SNR G349.7+0.2 is known to be interacting with nearby molecular clouds because of clear evidence: there are  five 1720 MHz OH masers along the radio emission ridge of the SNR (Frail et al. 1996); shock-excited near-infrared H$_{2}$ emission has been found toward the center of the remnant as well as OH absorption (1665 and 1667 MHz) in the remnant (Lazendic et al. 2010), and CO expansion caused by the SNR shock has been observed (Reynoso \& Mangum 2001). 
Because most 1720 MHz OH masers are seen as signposts of SNR-molecular cloud interaction (Wardle \& Yusef-Zadeh 2002), their velocities have been used to determine kinematic distances to associated SNRs based on a rotation curve model. 
The velocities of the 1720 MHz OH masers (14.3 to 16.9 km~s$^{-1}$) are nicely consistent with the velocity of shocked CO clouds (16.5 km~s$^{-1}$), also with the far 3 kpc arm velocity of $\sim$ 16.3 km~s$^{-1}$. 
Combined with the HI absorption analysis, we conclude that the 16.5 km~s$^{-1}$ CO
 clouds are behind but shocked by SNR G349.7+0.2 and this SNR-cloud interaction excites the near-infrared H$_{2}$ emission and the 1720 MHz OH maser emission in the region.     

An HI 21-cm absorption line study of G347.9+0.2 has previously been done using the Parkes hydrogen line interferometer (Caswell et al. 1975). 
Our absorption spectrum confirms the previous major absorption features near -110 km~s$^{-1}$, -62.5 km~s$^{-1}$, and 6 km~s$^{-1}$. 
But we find more fine structure because of the better quality of the SGPS data (higher continuum sensitivity and HI spectral resolution) and the improved methods. 
We find new absorption features at -35 and -189 km~s$^{-1}$, no absorption at 20 and 38 km~s$^{-1}$, much lower noise for baselines above 20 km~s$^{-1}$ and below -125 km~s$^{-1}$.  
Caswell et al. (1975) explained the -62.5 km~s$^{-1}$ feature as from the far ring of 3-4 kpc arm and the 6 km~s$^{-1}$ feature as from local gas, based on 1970s knowledge of the Galactic center, so they suggested a distance of 13.7 kpc $\le$ d $\le$ 23 kpc for SNR G349.7-0.2. 
However, our analysis utilizes new information that 
 the near and the far 3 kpc arms have kinematic velocities of -110 and -16.5 km~s$^{-1}$ towards G347.9+0.2 respectively, combined with our new result of reliable absorption at 6$\pm$7 km~s$^{-1}$. 
Thus we find the SNR has no absorption associated with the 16.5 km~s$^{-1}$ clouds in the far 3 kpc arm. 
 In this case, the -62.5 km~s$^{-1}$ feature is likely caused by much closer HI clouds whose motion follows the normal rotation curve model. 
 Our conclusion is strongly supported by other evidence: 
 the HI absorption at -62.5 km/s is also clearly seen in absorption spectra of the nearby SNRs G348.5+0.1 and G348.5-0.0, which both have distances of less than 9.5 kpc (Tian \& Leahy 2012);  
and the remnant is interacting with 16.5 km/s clouds in the far 3 kpc arm. 
 Previously, the 1720 MHz OH masers at 14.3 to 16.9 km~s$^{-1}$  were assumed to
 be outside of the Solar circle based on the circular rotation curve model, 
 so  SNR G349.7+0.2 was suggested to have a kinematic distance of $\sim$ 22 kpc. 
 Now we know that the circular rotation curve model is not valid near the GC, so we conclude that the SNR and its nearby clouds are located in the far 3 kpc arm, well within the Solar circle. 

SNR G349.7+0.2 is located at the near edge of the far 3 kpc arm because there 
is no absorption from the main part of the far 3 kpc arm (which has center 
velocity 16.5 km~s$^{-1}$). The far 3 kpc arm is at a distance of $\sim$ 11.5 kpc (Dame \& Thaddeus 2008, Jones et al. 2013). No precise thickness of the 3 kpc arm scale is available so far, so we here give an approximate distance of $\sim$ 11.5 kpc for SNR G349.7-0.2.  

Previous CHANDRA X-ray observations of SNR G349.7+0.3 gave a column density of 7 $\times$ 10$^{22}$ cm$^{-2}$. 
This is consistent with the estimate from adding HI (2.8 $\times$ 10$^{22}$ cm$^{-2}$, assuming T$_{s}$=140 K) and H$_2$ (4.6 $\times$ 10$^{22}$ cm$^{-2}$, assuming CO-to-H$_2$ conversion factor $\it{X}$ = 1.8$\times$10$^{20} $cm$^{-2}$ K$^{-1}$ km~s$^{-1}$) emission (Lazendic et al. 2005). 
Our calculation of the column density against G349.7+0.2 by HI absorption (using T${s}$=100 K) is $\sim$1.4 x 10$^{22}$ cm$^{-2}$), similar to Lazendic et al. (2005). 
The HI column density measures neutral hydrogen, whereas the X-ray column density is sensitive to the heavier element component in the total gas (HI, H$_{2}$ and HII) plus dust. 
So the X-ray column density could easily be 3 or more times higher than the HI column density.  
In addition, it is generally known that the column density value is strongly related with direction of a source, because the Galactic HI has a different distribution along different lines of sight. 
For example, SNR G18.8+0.3 has a distance of 12 kpc and column densities of 2 x 10$^{22}$ cm$^{-2}$ from  HI and 3.2 10$^{22}$ cm$^{-2}$ from H$_{2}$) (Tian et al. 2007).  
Generally, there are more clouds towards the GC than other directions, so there is no inconsistency in column densities with G349.7+0.2 being at 11.5 kpc.

It should be noted that kinematic distances to Galactic objects always include uncertainty due to uncertain parameters of different rotation curve models and wide-spread non-circular streaming motions. 
This uncertainty may reach up to 20\% 
at the particular longitude of some objects (Jones et al. 2013). 
SNR G349.7+0.2 is within the far 3 kpc arm: its distance's uncertainty is from the R$_{o}$ value and the width of the arm. 
The IAU standard is R$_{o}$ = 8.5 kpc, but recent measurements show a smaller value, e.g. 8.4$\pm$0.6 kpc from VLBI trigonometric parallax measurements (Reid et al. 2009), 8.05 $\pm$ 0.45 kpc from VLBI astrometry (Honma et al. 2013), 8.33$\pm$0.19 kpc from optical and near-infrared observation of known RR Lyrae stars in the bulge (Dekany et al. 2013) etc. 
The uncertainty of R$_{o}$ is about 0.5 kpc. 
The width of the far 3kpc arm is not yet known but should be smaller than the error of R$_{o}$. We therefore estimate a distance of 11.5$\pm$0.7 kpc for the far 3 kpc arm.         

In summary, we significantly revise the distance to G349.7+0.2 from a previous 22 kpc to 11.5 kpc, based on better understanding of the GC structure. 
We previously revised distances to three other SNRs nearby G349.7+0.2 (Tian \& Leahy 2012). 
One should use caution when considering old kinematic distances of Galactic objects (e.g. SNRs, pulsars and HII regions) in the range of -12$^{o} \le l \le 12^{o}$ and having distance estimates of $\ge$ 5.5 kpc.

\subsection{Luminosity, age, density and explosion energy of G349.7+0.2}
G349.7+0.2 has been suggested as one of the brightest Galactic sources in the radio (Shaver et al. 1985), X-ray (Slane et al. 2002) and GeV $\gamma$-ray wavebands (Castro \&  Slane 2010) because of its large distance. 
We give its luminosity based on the new distance measurement here: L$_{1GHz}\sim 3.2\times 10^{17}$ Watt Hz$^{-1}$, L$_{X(0.5-10.0keV)}\sim 1.0 \times 10^{37}$ erg s$^{-1}$, L$_{\gamma(0.1-100GeV)}$ $\sim$3.6 $\times$ 10$^{34}$ erg s$^{-1}$.   

The angular diameter of G349.7+0.2 ($\sim$2$^{\prime}$) from Chandra (Lazendic et al. 2005), yields a radius of R = 3.3 pc (d = 11.5 kpc) for the SNR.  
Lazendic et al.(2005) find a shock velocity of $\sim$710 km s$^{-1}$ based on the X-ray measured plasma temperature. 
The small remnant size and fast shock velocity indicate that the remnant is still evolving in the Sedov phase. 
So we estimate its age of $\sim$1800 yr by applying the Sedov model (Cox 1972, also see equations 4 and 5 of Reynoso \& Mangum 2001) using the known remnant radius and shock velocity. 
The SN explosion energy depends on the environment density n$_{0}$. 
Using the X-ray emission measure from the Chandra spectrum of 9.9$\times$10$^{59}$ cm$^{-3}$ (Lazendic et al. 2005) and a Sedov interior density profile (see Leahy et al. 2013), 
the ISM density is n$_{0}$ $\sim$ 10 cm$^{-3}$ (d = 11.5kpc). 
G349.7+0.2 is then found to have a low explosion energy of 2.5$\times$10$^{50}$ ergs.

\begin{acknowledgements}
We thank the anonymous referee for his/her constructive comments. WWT acknowledges support from the China Ministry of Science and Technology, the NSFC, and the CAS. DAL s grateful for the Natural Sciences and Engineering Research Council of Canada.  This publication is partly supported by a grant from the John Templeton Foundation and NAOCAS.   
\end{acknowledgements}

\end{document}